\documentclass[journal]{IEEEtran}
\ifCLASSINFOpdf
\else
\fi
\usepackage{graphicx}
\usepackage[table,xcdraw]{xcolor}
\usepackage[TABBOTCAP]{subfigure}
\usepackage{bm}
\usepackage{upgreek} 
\usepackage{amsmath}
\usepackage{breqn}
\usepackage{color}
\usepackage{cite}
\usepackage[none]{hyphenat}
\usepackage{algorithm}
\usepackage{algorithmic}
\usepackage{amssymb}

\usepackage{multirow}

\usepackage{float}
\usepackage{caption}
\usepackage{array}
\usepackage{booktabs}
\usepackage[colorlinks,
linkcolor=blue,
anchorcolor=blue,
citecolor=blue]{hyperref} 
\usepackage{amsmath}

\usepackage{color,xcolor}
\usepackage{makecell}

\usepackage{xr}
\usepackage{threeparttable}
\usepackage{footnote}

\begin{document}
%
\title{BAN: Neuroanatomical Aligning in Auditory Recognition between Artificial Neural Network and Human Cortex}
%
%
%

\author{Haidong~Wang$^*$,~Pengfei~Xiao,Ao~Liu, Jianhua~Zhang,~and Qia~Shan
	\thanks{H. Wang, P.Xiao, A.Liu, J. Zhang and Q. Shan are from Hunan University Of Technology and Business, Changsha 410082, China (e-mail: whd@hutb.edu.cn; 2893666867@qq.com; 2496556459@qq.com; zhangjianhua6682@126.com; s540534349@163.com).}
}

%
%

\markboth{Journal of \LaTeX\ Class Files,~Vol.~14, No.~8, August~2015}%
{Shell \MakeLowercase{\textit{et al.}}: Bare Demo of IEEEtran.cls for IEEE Journals}
%



\maketitle

\begin{abstract}
Drawing inspiration from neurosciences, artificial neural networks (ANNs) have evolved from shallow architectures to highly complex, deep structures, yielding exceptional performance in auditory recognition tasks.
However, traditional ANNs often struggle to align with brain regions due to their excessive depth and lack of biologically realistic features, like recurrent connection.
To address this, a brain-like auditory network (BAN) is introduced, which incorporates four neuroanatomically mapped areas and recurrent connection, guided by a novel metric called the brain-like auditory score (BAS).
BAS serves as a benchmark for evaluating the similarity between BAN and human auditory recognition pathway.
We further propose that specific areas in the cerebral cortex, mainly the middle and medial superior temporal (T2/T3) areas, correspond to the designed network structure, drawing parallels with the brain's auditory perception pathway.
Our findings suggest that the neuroanatomical similarity in the cortex and auditory classification abilities of the ANN are well-aligned. 
In addition to delivering excellent performance on a music genre classification task, the BAN demonstrates a high BAS score.
In conclusion, this study presents BAN as a recurrent, brain-inspired ANN, representing the first model that mirrors the cortical pathway of auditory recognition.
\end{abstract}


%
\IEEEpeerreviewmaketitle

\begin{figure*}
	\centering
	\includegraphics[width=6.7in]{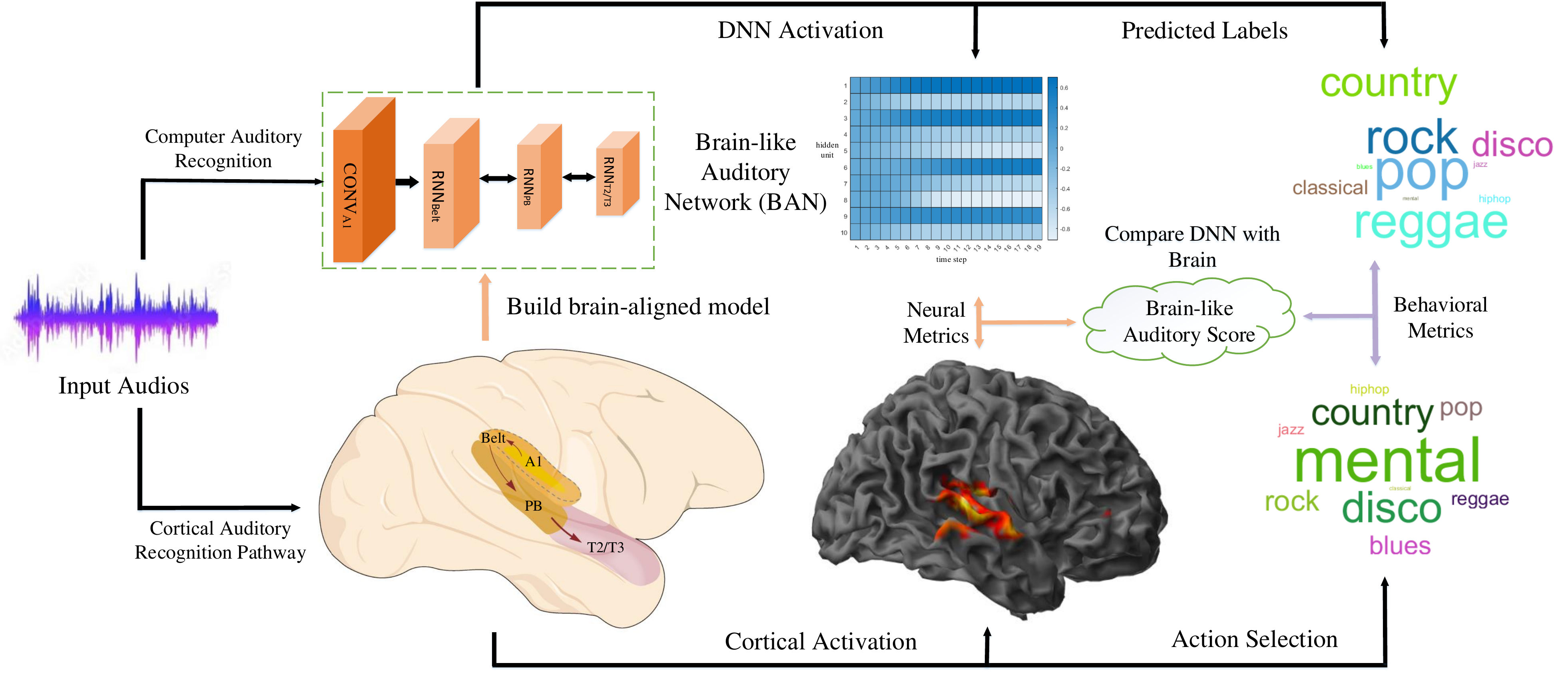}
	\caption{
		\textbf{
		Collaborative model between ANNs and neuroanatomy with brain-like auditory score (BAS).}
		Using the quantitative BAS, we draw inspiration from brain and apply this idea to inform the model of BAN. 
		BAN consists of four regions mapped to the primary auditory cortex (A1), the belt or peripheral area (Belt), the parabelt cortex (PB), and the middle and superior temporal cortex areas (T2/T3). 
		The CONV\textsubscript{V1} layer is a traditional convolutional layer responsible for preprocessing and data size reduction. 
		The RNN\textsubscript{Belt} and RNN\textsubscript{PB} and RNN\textsubscript{T2/T3} reference recurrent neural networks for modeling\cite{park2020circuit}, as detailed in Sec.~\ref{sec:cornet_s_def}. 
		The upper-right image illustrates the auditory recognition activations and predicted genre labels in the brain-aligned neural network, while the lower-right figure shows human activations and choices for the same stimulus on the far left. 
		This comparison highlights the relationship between BAN's prediction performance and brain auditory recognition responses, as explained in Sec.\ref{sec:experiments}.
	}
	\label{fig:introduction}
\end{figure*}

\section{Introduction}
%
%
%
%


\IEEEPARstart{H}{earing} plays a vital role in human sound recognition and is especially important for the comprehension and creation of music.
One key task in this domain is genre classification, which involves predicting the genre of a piece of music based on its audio signal.
Music genres, such as jazz, rock, and classical, serve as descriptive labels that provide high-level information about a musical piece.
As noted by previous work~\cite{tzanetakis2002musical}, genres are classes introduced by humans to categorize musical works. 
These genres are defined by shared characteristics among the music that belongs to them.
These characteristics are generally linked to specific musical instruments, rhythmic patterns, and harmonic structures.
Automating music genre classification can assist or even replace human effort in this process, making it a valuable tool in music information retrieval systems.
As Nam~\cite{nam2018deep} highlights, music genre recognition plays a important role in auditory recognition and recommendation algorithms for music streaming services.
In solving the tasks of music classification and tagging, it is essential for the computer to recognize and identify musical patterns.
The traditional framework's features are key to designing the pipeline, involving both feature engineering and classifier development.

When humans distinguish music, they rely on features such as the sound of instruments or rhythm. 
Teaching machines to recognize these features has traditionally involved hand-engineering using domain-specific knowledge\cite{nakai2018encoding}.
However, with the rise of deep learning, new approaches and improved performance have emerged. 
Deep learning methodologies reduce the need for extensive domain knowledge by automatically extracting features. 
Typically, deep learning models build pipelines through linear transformations, nonlinear activation functions, and optional pooling operations\cite{thompson2021training}. 
As data passes through these layers, the model learns to extract features, enabling end-to-end learning.

Deep learning has also made significant strides in modeling neural mechanisms within the neuroscience community~\cite{kubilius2019brain-like}. 
Remarkably, in artificial neural networks (ANNs) trained for image classification ~\cite{Deng2009ImageNet}, the middle layers can partially explain why neurons in the visual cortex's middle layer respond to specific image features~\cite{yamins2014performance,suk2012novel,gucclu2015deep,murugesan2017brain,hu2021neuroscience,wang2021memristive}. 
In addition, these networks can predict primate image classification behavior and cortical activations to some extent~\cite{rajalingham2018large,kubilius2016deep}. 
Such brain-inspired models open up exciting possibilities to brain-computer interfaces, where these modles can infer expected responses along human cortical pathway~\cite{bashivan2019neural}.

To more accurately model brain processing, relying solely on traditional visual datasets for architectural improvements is becoming increasingly impractical. 
While deeper ANNs have significantly improved performance~\cite{ILSVRC15,luo2021trajectory}, raising brain-like scores remains a challenge~\cite{rajalingham2018large,su2020incremental}. 
Additionally, in earlier stages, only certain modules could be clearly mapped to specific areas of the visual pathway, and these connections between a few visual areas and numerous complex modules in architectures like GoogleNet~\cite{szegedy2015going} or other deep network are not well-defined. 
Furthermore, as networks for audio recognition tasks achieve higher accuracy, they tend to become progressively deeper, with only a few exceptions where brain-inspired architectures are used for audio recognition~\cite{kim2024spontaneous}.

To deal with the interpretability challenges of auditory recognition models, we propose that aligning artificial neural networks (ANNs) with neuroanatomy can produce more interpretable, shallower, and brain-like neural networks, which we term the brain-like auditory recognition network (BAN). 
BAN is a recurrent and shallow architecture that mimics the auditory cortical pathway, giving it a structure more closely aligned with neuroanatomy. 
In Sec.\ref{sec:data_analysis}, we introduce a method to analyze cortical data and human behavior during auditory recognition. 
In Sec.\ref{sec:bas}, we develop an innovative measurement approach for predicting neural activations and human auditory recognition choice, referred to as the brain-like auditory score (BAS).

Our researches involve a novel benchmark consisting of cortical and behavioral recordings, where BAN demonstrates excellent performance in evaluating the similarity between the model and the auditory recognition pathway, while also achieving strong results on a music genre classification task~\cite{nakai2022music}. 
These outcomes are largely attributed to BAN's brain-like structure, consistent with existing knowledge of the human cortical pathway's response to audio~\cite{TangSchrimpfLotter2018Recurrent, yin2020deep, kar2019evidence}. 
Lastly, in comparing the activations in BAN's RNN$_{\text{T2/T3}}$ layers with responses in human middle and superior temporal regions (T2/T3), we find BAN accurately infers cortical responses. 
This is also the first auditory recognition network to achieve this based on neural activations.


\section{Related Works} 

\subsection{Brain-like Neural Network}\par

As research in computer vision and human vision advances, there has been significant growth in brain-like vision studies. 
Compared to models of the brain’s ventral stream, which have seen extensive development, image-computable models of the dorsal stream have been relatively scarce. 
Previous efforts to model the dorsal stream have involved training deep neural networks to detect motion \cite{rideaux2020but} or classify actions \cite{gucclu2017increasingly} using video inputs. 
However, these models fail to fully align with the neuroscientific understanding that the dorsal stream is responsible for object localization and action guidance, often referred to as the ``where" or ``how" visual pathway. 
More recent work focused on training a dorsal-stream model to mimic human head movements during visual exploration\cite{mineault2021your}. 
Additionally, the predictive learning is employed to train parallel pathways, leading to the emergent development of ventral-like and dorsal-like representations as a result of structural segregation\cite{bakhtiari2021functional}.

In addition to brain-like vision, numerous studies have explored brain-like approaches in other domains. 
These methods have successfully linked the structure of neural activity to computational functions in areas such as audition\cite{kell2018task}, olfaction \cite{wang2021evolving,singh2023emergent}, thermosensation\cite{haesemeyer2019convergent}, perceptual discrimination\cite{mante2013context}, facial recognition\cite{higgins2021unsupervised}, and navigation\cite{banino2018vector,cueva2018emergence}. 
Additionally, pioneering efforts have demonstrated the ability of simple brain-inspired controllers to replicate animal locomotion\cite{ijspeert2007swimming,grillner2007modeling,J2020Reproducing}, illustrated how biomechanics can shape neural representations of movement\cite{2013Preference}, and uncovered similarities between the representation of movement in artificial and biological neural networks\cite{2000Direct,2020A,2015A}. 
Our study leverages brain-like mechanisms, implemented through a recurrent network, to effectively extract audio features from multiple music fragments. 
Furthermore, we aim to design a brain-like network that not only achieves a high brain-like auditory score (BAS) but also surpasses existing recognition models on the Music Genres dataset~\cite{nakai2022music}.

\par

\subsection{Auditory Recognition}

Recent researches utilizing artificial neural networks (ANNs) have shed light on the principles based on the development of sensory functions in cortex\cite{richards2019deep,hassabis2017neuroscience,kell2019deep,saxe2021if}. 
It has been suggested that brain-like sensory encoding can emerge as a by-product of optimizing ANNs to process natural stimuli.
For instance, ANN models trained to classify natural images have been shown to predicte visual cortical activations and even influence real neuron responses beyond their natural limits\cite{cadieu2014deep,yamins2014performance,bashivan2018neural}. 
Similarly, a trained ANN to classify music was able to mimic human auditory cortical activations\cite{kell2018task}, suggesting that task-specific learning could serve as a efficient method for modeling auditory cortex functions. 
Additionally, research has explored how music-selectivity in neural circuits might develop as a by-product of adapting to natural audio processing\cite{hauser2003evolution,trainor2015origins,honing2015without,mlynarski2019ecological}, with the statistical patterns of natural sounds potentially shaping the brain’s innate musical foundation. 
To further align brain-like models with human intelligence and neurosciences, the brain-like score has been introduced as an integrated benchmark for evaluating such models~\cite{schrimpf2020integrative}.

In this work, we fully leverage the latest advancements in audio perception and music classification techniques. 
By applying these approches,  functional magnetic resonance imaging (fMRI) is processed from the Music Genres dataset ~\cite{nakai2022music} to identify cortical regions involved in music genre perception. 
Furthermore, we developed the BAS to assess the similarity between BAN activations and brain activations, providing a benchmark for evaluating their alignment.

\section{BAN: Brain-like Auditory Network}

In this part, we present the proposed BAN with three steps. 
Firstly, the design principles and aim behind our approach is outlined. 
Secondly, a detailed explanation of each component within the BAN pipeline is provided. 
Finally, the loss function used to train the BAN is introduced.

\subsection{Design Criteria}

We designed the BAN based on the following two criteria~\cite{kubilius2018predict}:

(1) \textbf{Architecture}: Among neural networks with same recognition performance, we prioritize brain-like networks due to their interpretability and their ability to align with anatomical constraints. 
We utilize ANNs as their neurons serve as fundamental units of information processing, and all neural responses in ANNs can be directly mapped to cortical activations ~\cite{yamins2016using}. 
Additionally, recurrent connections are naturally incorporated for auditory recognition because of the temporal nature of audio sequences. 
Since responses in auditory ventral pathway also exhibit temporal characteristics, the BAN is designed to generate activations over time.

(2) \textbf{Predictivity}: 
The intermediate layers and final outputs of the model that aligns with neuroanatomical constraints (neural responses) exhibit more accurate behavior. 
It enables the proposed neural network to effectively infer both brain activations and category choice.

Our goal is to achieve a high level of auditory recognition similarity between ANN and human brain. 
Moreover, auditory recognition in the human brain is reflected through human decision-making. 
In the auditory recognition pathway, the primary auditory cortex (A1) serves as the initial stage for preprocessing input signals, while the Belt and PB regions integrate sound signals across spatial dimensions, and the T2/T3 regions generate predictive auditory labels~\cite{b11,b13,b14}.
We now introduce the pipeline for BAN.

\begin{figure*}
	\centering
	\includegraphics[width=5.7in]{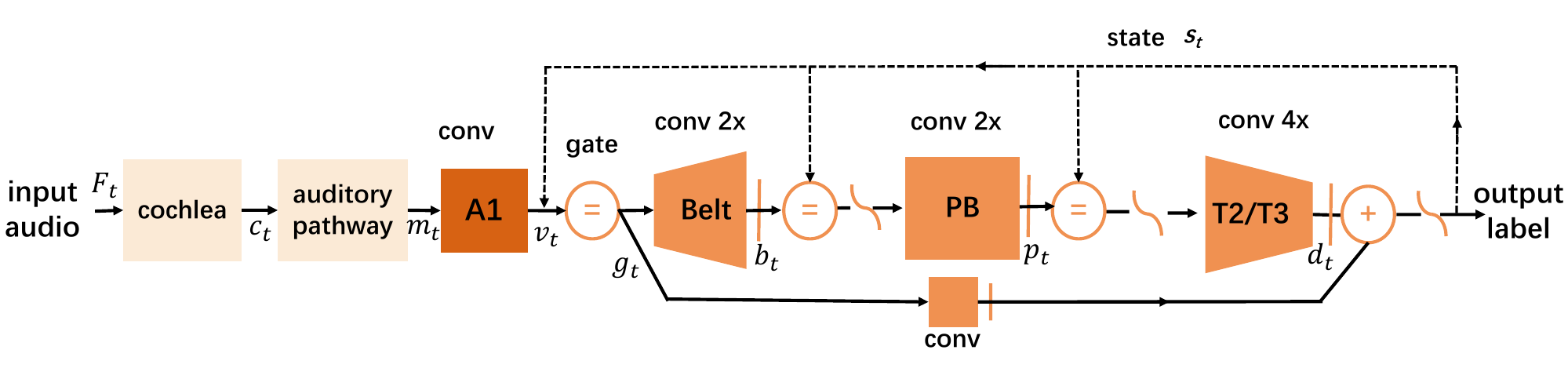}
	\caption{
		\textbf{
			The BAN circuitry is designed based on the human auditory cortex pathway.}
		Key cortical areas involved in auditory recognition are highlighted with orange modules. 
		Auditory neural signals are generated from the input music $F_t$ via the cochlear hair cells. 
		In the ventral pathway, A1 processes auditory representations $v_t$, and neurons in the ventral ``what" pathway (including A1, Belt, PB, and T2/T3) extract auditory features from the preprocessed neural signal $m_t$. 
		The Belt and PB are particularly important for refining features based on the temporal and rate codes $v_t$, and the output $p_t$ from PB is passed to the next module. 
		The Belt, PB, and T2/T3 regions are recurrent. 
		Finally, fully connected layers serve as the coder to generate auditory labels. 
		Solid arrows show feature flow within a single time step, while dashed black arrows indicate temporal connections.
	}
	\label{fig:pipeline}
\end{figure*}

\subsection{BAN Pipeline} \label{sec:cornet_s_def}

Inspired by the brain’s auditory recognition pathway, we establish a neuroanatomical mapping between cortical areas and ANN layers, as illustrated in Fig.~\ref{fig:introduction}. 
To facilitate neural network comparison, this mapping by identifying the ANN layer that best corresponds to activations in a specific cortical region is created. 
Ideally, these responses are inferred by the neural network without requiring unnecessary parameters. 
BAN consists of four parts: a convolutional layer and three recurrent layers, which correspond to the auditory recognition pathway regions A1, Belt, PB, and T2/T3. 
Additionally, the recognition predictor converts the output of T2/T3 into action selection activations, as shown in Fig.~\ref{fig:pipeline}. 
This straightforward approach of explicitly segmenting brain regions is a key step in designing a brain-like auditory recognition model. 
We are also focused on discovering more generalized structures, such as a unified neural network model without distinct cortical regions or varied connections that could enhance BAS in future research.

As shown in Fig.~\ref{fig:pipeline}, we compute a decoding $c_t$ from the cochlea field based on the current input $F_t$, representing the ventral pathway responsible for extracting audio features. 
The auditory pathway selects a view $m_t$ for the cochlea response $c_t$, and this output is then transmitted to the ventral ``what" cortical pathways.


The ventral pathway is responsible for recognizing sounds within the auditory field. 
The BAN (Fig.~\ref{fig:introduction}) seeks to compete with state of art methods on the BAS by converting very deep feedforward structures into a shallow and recurrent network. 
Besides, BAN is inspired by ResNets, which are among the best models on behavioral benchmarks\cite{nakai2022music} and can be viewed as unrolled recurrent structure. 
Recent research has also shown that sharing weight in recurrent neural networks is feasible without significantly compromising performance\cite{liao2016bridging}.

Moreover, BAN is explicitly designed with an anatomical mapping to cortical regions. 
While model comparison establish this mapping by identifying the layer that best corresponds to neural activations in a specific cortical region, BAN aims to provide this mapping inherently, without the need for additional parameters. 
BAN features four computational areas, modeled after the auditory regions A1, Belt, PB, and T2/T3, along with a linear class decoder that translates neuron activity in the final auditory area into behavioral decisions. 
This initial assumption of distinct areas with recurrent circuitry was our first step toward creating a shallow model. 

Next, we provide a detailed explanation of each component within the brain-like auditory recognition architecture.

\subsubsection{Primary Auditory Modules}
Typically, processed auditory signals reach A1 in the cortical temporal lobe, which is essential for processing fundamental sound features like pitch and volume. 
The input is a data matrix represented by a time-frequency representation $ m_t $. 
In our work, this input is a Mel-spectrogram.


Each auditory area is designed with specific neural circuitry where neurons perform fundamental computations, such as convolution, addition, normalization, nonlinearity, or pooling. 
The circuitry remains consistent across all auditory areas (except for A1$_{\text{COR}}$), though the total number of neurons varies in each region. 
To manage the high computational costs, the first region, A1$_{\text{COR}}$, applies a 7 $\times$ 7 convolution with a stride of 2, followed by 3 $\times$ 3 max pooling, and then a 3 $\times$ 3 convolution.

%

\subsubsection{Belt and PB Modules}

The Belt and Parabelt (PB) areas are key regions within the auditory cortex, primarily responsible for processing complex auditory information. 
The RNN$_{\text{Belt}}$, surrounding Conv$_{\text{A1}}$, represents the first level of auditory cortex beyond Conv$_{\text{A1}}$ and receives direct inputs $ v_t $ from the primary auditory cortex. 
As a secondary auditory processing region, RNNBelt is essential for analyzing more complex sound features than Conv$_{\text{A1}}$, integrating information from Conv$_{\text{A1}}$ to provide a more refined understanding of sounds, such as recognizing intricate patterns and spatial localization of sound sources.

RNN$_{\text{PB}}$, located adjacent to RNN$_{\text{Belt}}$, represents an even higher level of auditory processing. 
It receives inputs $ b_t $ from RNN$_{\text{Belt}}$ and connects with various brain regions involved in memory, emotion, and multisensory integration. 
RNN$_{\text{PB}}$ plays a crucial role in advanced auditory scene analysis, such as distinguishing multiple voices in a noisy environment or interpreting modulated sounds like music and speech. 
It also integrates auditory information with other sensory inputs, helping create a unified perception of the surrounding environment.



\subsubsection{Middle and Superior Temporal Modules (T2/T3)}

The RNN$_{\text{T2/T3}}$ cortices are crucial brain regions involved in multiple aspects of sensory processing, particularly auditory recognition. 
Located in the temporal lobe, these areas are essential for interpreting and comprehending complex auditory signals, such as music.

T2 is primarily known for its role in auditory processing, particularly in perceiving motion and integrating audio-visual information. 
In auditory recognition, the T2 cortex becomes active when visual cues need to be integrated with auditory signals. 
As shown in TABLE~\ref{tab:prediction_accuracies}, T3, particularly the superior temporal gyrus (STG), plays a direct role in auditory processing and is essential for recognizing complex sounds, such as music. 
Within the STG, Heschl's gyrus (HG) is the first brain area to receive input audio, while surrounding regions, including the planum temporale (PT), are involved in processing higher-order sound features like language comprehension. 
The posterior part of the STG and the nearby superior temporal sulcus contribute to analyzing more complex sound attributes, such as prosody and the emotional content of speech. 
T3 also has extensive connections with other brain regions, supporting its role in integrating auditory information with other sensory modalities and cognitive functions.

In auditory recognition, these areas collaborate to decode and interpret sounds, enabling individuals to effectively recognize and respond to various auditory stimuli, such as distinguishing speech sounds, understanding spoken language, and appreciating music. 
The Belt$_{\text{COR}}$, PB$_{\text{COR}}$ and T2/T3$_{\text{COR}}$ areas each perform two 1 $\times$ 1 convolutions, followed by a bottleneck-style 3 $\times$ 3 convolution, which expands the feature set fourfold, and concludes with another 1 $\times$ 1 convolution. 
Recurrence is implemented by passing the output of an region through that same region multiple times. 
For example, as shown by the ``gate" in Fig.~\ref{fig:introduction}, after Belt$_{\text{COR}}$ processes the input feature once, the result is reprocessed by Belt$_{\text{COR}}$ as new input. 
As depicted in Fig.~\ref{fig:structure_analysis}, Belt$_{\text{COR}}$ and PB$_{\text{COR}}$ are repeated twice respectively, while T2/T3$_{\text{COR}} $ is repeated four times, as this configuration was found to produce the best model performance with the fewest layers, according to our scores. 
Similar to ResNet, a convolution module is followed by batch normalization\cite{ioffe2015batch} and ReLU. 
In the current definition of BAN, there are no bypass or feedback connections across areas, and cochlea and auditory pathway processing are not explicitly modeled.

\subsubsection{Classification Decoder}

The decoder in BAN uses a straightforward linear classifier, which consists of weighted sums, with one sum for every object label. 
To minimize the number of neural activations feeding into the classifier, the responses for each feature map is averaged.

\subsection{Training Loss} \label{sec:loss}

The proposed BAN is trained by optimizing a combination of losses: a recognition loss and an auxiliary loss. 
The total BAN loss $L_{b}$ is defined as follows:
\begin{equation}
	L_{b} = L_r + L_u,
\end{equation}
where $ L_r $ is recognition loss depicted in Equ.~\ref{equ:recognition_loss}, $ L_u $ is auxiliary loss described in Equ.~\ref{equ:auxiliary_loss}.

\subsubsection{Recognition Loss}

To facilitate music recognition, we design the recognition loss to measure the difference between the output of BAN and the true label. 
This recognition loss $ L_r $ is defined as:
\begin{equation} \label{equ:recognition_loss}
	L_r = \frac{1}{N} 
		\sum_{i=1}^{N}(
			-(d_i * log(p_i) + 
			(1-d_i) * log(1-p_i))
		),
\end{equation}
where $ L_r $ the cross-entropy loss between the ground truth $d_i$ and the predicted label $p_i$, $ N $ is the number of samples.

\subsubsection{Auxiliary Loss}

L2 regularization is applied to both the dynamic parameter $\phi_t(s_t)$ and the model parameter $\theta$ to ensure proper regularization.
\begin{equation} \label{equ:auxiliary_loss}
	L_u = \frac{1}{2} \left\vert \left\vert \phi_t \right\vert \right\vert _2 ^2 
	+ \frac{1}{2} \left\vert \left\vert \theta \right\vert \right\vert _2 ^2.
\end{equation}

\section{BAS: Brain-like Auditory Score} \label{sec:bas}
In this section, the BAS metrics is introduced, which assess the similarity between the ANN and brain. 
BAS is a measurement evaluated on given data, incorporating both cortical and behavioral metrics.

To get quantitative measurement for cortical similarity, we reference the open-source platform Brain-Score~\cite{SchrimpfKubilius2018BrainScore} and introduce the BAS.
It is a measurement that measure the performance to predict 
(a) the average behavioral choices when listening to target music clips from the Music Genre dataset~\cite{nakai2022music}, and 
(b) the average cortical activation at each brain region in response to the same music clips in the human auditory areas from the Music Genre neuroimaging dataset~\cite{nakai2022music}. 
To provide a unified evaluation of BAN, we calculate the average of both behavioral and cortical measurement.

\subsection{Cortical Metrics}
\label{sec:neural-pred}

Cortical metrics is utilized, which capture responses in regions of brain, such as activations in T2/T3, to assess how accurately BAN predicts the audio in ANN~\cite{yamins2014performance}. 
This measurement requires two sets of audio in the format of $\text{audio} \times \text{neuroids}$, where neuroids represent either model interlayer responses.

A total of 540 audio from 10 different music were randomly shown to 5 subjects, and neural activations were recorded in the T2/T3 regions. 
Additionally, we identify and present the most predictive layers or areas, RNN$_{\text{T2/T3}} $, within BAN model.

In our sutdy, relationships are established to map the ANN to cortical regions, and these relationships are used to infer neural activations to the given music clips. 
To speed up it, we compress the dimensionality of the activations to specific components using principal component analysis~\cite{2002Principal}. 
The responses in T2/T3 are used to learn these mappings. 
The final neural similarity score for the auditory recognition cortex is represented by the Pearson correlation coefficient $s_r$, calculated as follows:
\begin{equation} \label{equ:cortical_metrics}
	s_r= \frac{1}{N} \sum_{j=1}^{N}
	\frac{\sum_{i=1}^{n} (y_i-\bar{y}) (y_i^\prime - \bar{y}^\prime) }{\sqrt{\sum_{i=1}^{n} (y_i - \bar{y})^2 (y_i^\prime - \bar{y}^\prime)^2 }},
\end{equation}
where $y$ represents the actual neural activation, and $y^\prime$ is the model's predicted activation, $n$ is the dimension of corresponding layer in BAN, 
$ N $ is the number of music sample used in our experiment. 
$\bar{y}$ and $\bar{y}^\prime$ are the median values of the actual and predicted neural responses, respectively, across all data points.

\subsection{Behavioral metrics}

The goal of behavioral metrics is to assess the similarity between BAN's outputs and human behavior in music recognition tasks~\cite{nakai2022music}. 
In human listening experiments, participants provide a music genre label, so the behavioral model is represented as a categorical label in auditory recognition. 
The primary focus is to achieve human-like intelligence, not just auditory classification accuracy~\cite{schrimpf2020integrative}. 
BAN excels in behavioral similarity, accurately predicting both labels and neural activations. 
In contrast, while traditional ANNs may achieve excellent classification performance, they often fall short in delivering strong brain-like prediction performance.

As outlined in Sec.\ref{sec:behavioral_data_collection}, to compare brain activation patterns with behavioral performance, we gathered music genre choices from subjects through additional behavioral experiments. 
The corresponding output of the BAN is the predicted genre label. 
Thus, the music class predictivity, or behavioral score, is modeled as the similarity between the actual music genre choices made by the subjects and the predictions made by BAN. 
We calculate the overall behavioral metric $s_b$ across all audio sequences to determine the predictivity score,
\begin{equation} \label{equ:behavioral_metrics}
	s_b = 
	\frac{\sum_{i=1}^{M} TP_i}
	{\sum_{i=1}^{M} (TP_i + FP_i + FN_i)},
\end{equation}
where $ TP_i $ represents the number of true positives for music genres $ i $ (correct predictions), 
$ FP_i $ represents the number of false positives for music genres $ i $ (incorrectly predicted as genres $ i $), 
$ FN_i $ represents the number of false negatives for class $ i $ (instances of class $ i $ incorrectly predicted as another music genres),
$ M $ is the total number of genres.

\subsection{Overall score}

To assess the overall performance of the BAN, we use BAS, which combines both the behavioral metric and the T2/T3 cortical metrics. 
The BAS, denoted as $s_{a}$, is calculated as the average of these two scores.
\begin{equation} \label{equ:score_btn}
	s_{a} = \frac{1}{3} [\frac{min(m,n)}{max(m,n)} + s_r + s_b].
\end{equation}
where $ m $, $ n $ is the number of modules in cortical autitory recognition pathway and BAN, respectively.
The first item as a whole measures the similarity between cortical and BAN structures.
$ s_r $ is the similarity of cortical metrics depicted in Equ.~\ref{equ:cortical_metrics}, $ s_b $ is the similarity of behavioral metrics described in Equ.~\ref{equ:behavioral_metrics}.
The BAS is designed without normalization across different score magnitudes, as this could unfairly punishment with small variances. 
Instead, every score is treated equally to ensure fair significance in the overall BAS calculation.

\section{Experiments} \label{sec:experiments}

We validate the effectiveness of our designed BAN through three steps. 
Firstly, the datasets and provide details on the implementation of our model is introduced. 
Secondly, that BAN functions as a valid, brain-like auditory recognition model through circuitry analysis is shown. 
Thirdly, the model's classification accuracy and its representation of music genres in both BAN and the cortex is discussed.

\subsection{Datasets} \label{sec:datasets}

We used the GTZAN dataset, which is one of the most widely utilized in music genre recognition tasks\cite{nakai2022music}. 
The dataset consists of 30-second audio files spanning 10 different genres: blues, reggae, classical, rock, country, disco, jazz, pop, metal, and hip-hop. 
From the original collection, we randomly selected 54 music pieces from each genre, resulting in a total of 540 music pieces for the study. 
All clips were normalized based on their root mean square values.

The fMRI experiment included 12 training runs and 6 test runs, for a total of 18 runs. 
Each run lasted 10 minutes and consisted of 40 music clips. 
In the training phase, 480 music clips were used, while the remaining 60 clips were reserved for the test runs. 
During each test run, a set of 10 music clips was shown four times in the same order. 
There is no repetition of clips in the training runs.

\begin{table}[h]
	\centering
	\footnotesize
	\caption{General Information of fMRI for participants.}
	
	\label{tab:fMRI_runs}
	
	\begin{tabular}{p{2.9cm}<{\centering}p{0.6cm}<{\centering}p{0.6cm}<{\centering}p{0.7cm}<{\centering}p{0.7cm}<{\centering}p{0.7cm}<{\centering}}
		\toprule[1.0pt]
		 & sub-01  & sub-02    & sub-03  & sub-04 & sub-05  \\ \midrule
		Sex     		  & M  &  M	   & M  & F  & F		  \\
		Age     		  & 30 			   & 30 	   & 23  & 33  & 25		  \\
		Music experience (years)     		  & 12  &  12	   & 4  & 15  & 10		  \\
		test samples (volumes) & 1350  & 1350 	   & 1350  & 1350  & 1350		  \\
		train samples (volumes) & 4050  & 4050 	   & 4050  & 4050  & 4050		  \\
		Total fMRI runs     		  & 18  & 18 	   & 18  & 18  & 18		  \\
		\bottomrule[1.0pt]
	\end{tabular}
	
\end{table}

\subsection{Human Data Analysis} \label{sec:data_analysis}

To assess the similarity between the ANN and human auditory recognition, we extract task-related activations and then identify the corresponding brain regions for comparison.


\subsubsection{fMRI Data Processing}
Motion correction is applied to each run using the Statistical Parameter Mapping toolbox (SPM 12), with all volumes aligned to the first image for every participant. 
We remove Low-frequency drift with a filter using a 240-second window. 
To improve method accuracy, the activation for each voxel is normalized by subtracting the mean and variance. 
Cortical surfaces are identified using FreeSurfer~\cite{dale1999cortical,fischl1999cortical}, which registers the anatomical data with functional voxels. 
For analysis, only cortical voxels are used as targets, and for a participant, we focus on the voxels identified within the cortex.

\subsubsection{Genre Representation Region}

To obtain reliable estimates of human areas associated with music genres, we use the these step: 
Firstly, all activation recordings is randomly split into training sets (75\%) and test sets (25\%). 
Utilizing the optimal weights from the genre-label model\cite{nakai2021correspondence}, we fit an encoding model with genre-label features using the training data and assess model accuracy with the test sets. 
Parameter fitting is done with general linear model. 
The random resampling is executed 100 times, and voxels that show prediction accuracy in more than 75\% of repetitions are picked out for region of interest. 
As a result, we induce 473 voxels in the region of interest for participant, 468 for sub-01, 581 for sub-02, 1,593 for sub-03, and 529 for sub-05. 
And all subsequent analyses use these extracted region of interest.

\subsubsection{Behavioral Data Collection} \label{sec:behavioral_data_collection}

To verify that cortical activation in response to music was linked to behavioral accuracy in musci recognition, we utilized additional behavioral data\cite{nakai2022music}. 
As shown in TABLE~\ref{tab:fMRI_runs}, these experiments took place in a soundproof place with the same participants from fMRI study. 
Participants first listened to 3 original 30-second music for each genre, selected from the 460 clips randomly not utilized in fMRI test, as a reference. 
During this training session, participants were informed of the correct genres. 
They listened to the 60 audio utilized in fMRI experiment and classified each clip's genre by selecting one of 10 options on an answer sheet. 
Each audio was played once, in the same sequence as in the fMRI test.

\subsection{Implementation Details}
However, we determined that the available data was insufficient to effectively train the parameters of the CNN. 
As a result, we applied data augmentation techniques to enhance the dataset.
Data augmentation is the process of generating new synthetic training samples by applying small modifications to the original dataset. 
The goal is to make the model robust to these variations and improve its generalization ability. 
For this approach to be effective, the added perturbations must preserve the original label of the training sample. 
Common techniques include adding noise, shifting the audio start point, altering speed, and changing pitch.

The Mel-frequency cepstral coefficient (MFCC) model is utilized to extract various genres-related information from the dataset \cite{rabiner2010theory}. 
For timbral and loudness information, the frame duration is set to 25 ms, with a 50\% overlap between adjacency. 
For  tonal and rhythm information, the frame duration is 3 seconds, with a 33\% overlap, and each feature was averaged over 1.5 seconds. 
Additionally, the MIR toolbox was utilized to extract MFCC features \cite{lartillot2008matlab}. 

Each auditory cortical region is represented by a specific ANN that performs classical operations such as convolution. 
These modules correspond to audio cortical areas, and we adjust the number of neurons in each auditory brain area. 
Given the computational demands, we implemented the model using Python and the PyTorch framework~\cite{paszke2019pytorch}.


\subsection{Circuitry and Ablation Analysis}
\begin{figure}[t]
	\centering
	\includegraphics[width=\linewidth]{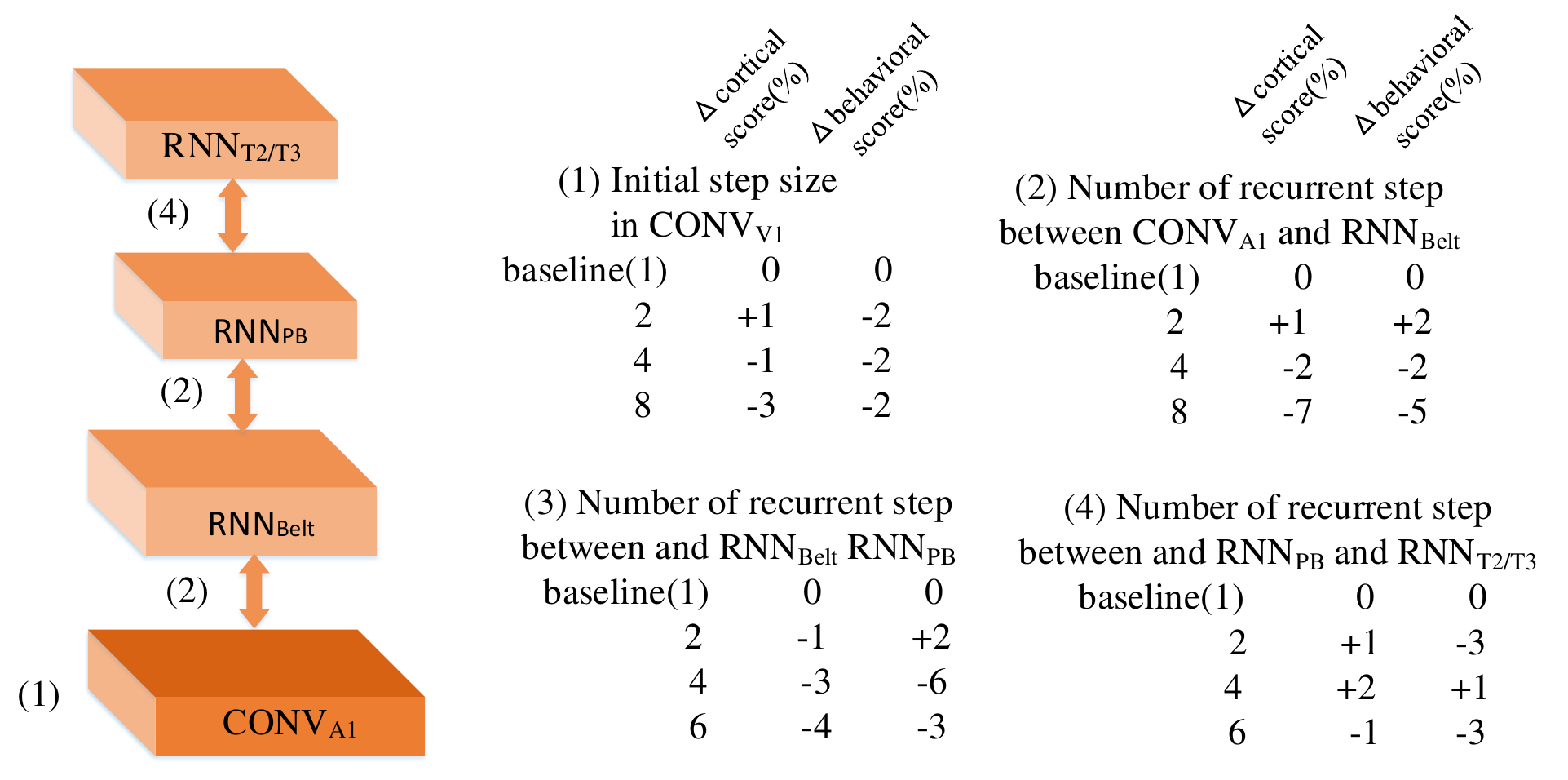}
	\caption{
		\textbf{
			BAN circuitry analysis.  
		}
		Each row shows how the highest accuracy on the GTZAN dataset and the BAS score change relative to the baseline model when a specific hyperparameter is modified, including cortical score $ s_r $ and behavioral score $ s_b $.
	}
	\label{fig:structure_analysis}
\end{figure}

While we believe that BAN offers a closer approximation to the anatomy of the auditory pathway compared to current ANNs—primarily due to its limited number of regions and inclusion of recurrence—it remains incomplete in several aspects. 
From a neuroscientific perspective, in addition to lacking biologically plausible learning methods, an ideal method of the auditory stream would incorporate more anatomical and circuitry-level details, such as the cochlea or the medial geniculate nucleus. 
Similarly, the addition of skip connections was not based on brain circuitry properties but rather adopted from complexity gradient\cite{he2016deep}  as a solution to the degradation problem in deep model. 
It's important to note that not all  structural choices are effective. 
We tested thousands of structures before identifying the BAN circuitry, as shown in Fig.~\ref{fig:structure_analysis}.

\subsection{Classification Accuracy based on BAN}
\begin{figure}[t]
	\centering
	\includegraphics[width=\linewidth]{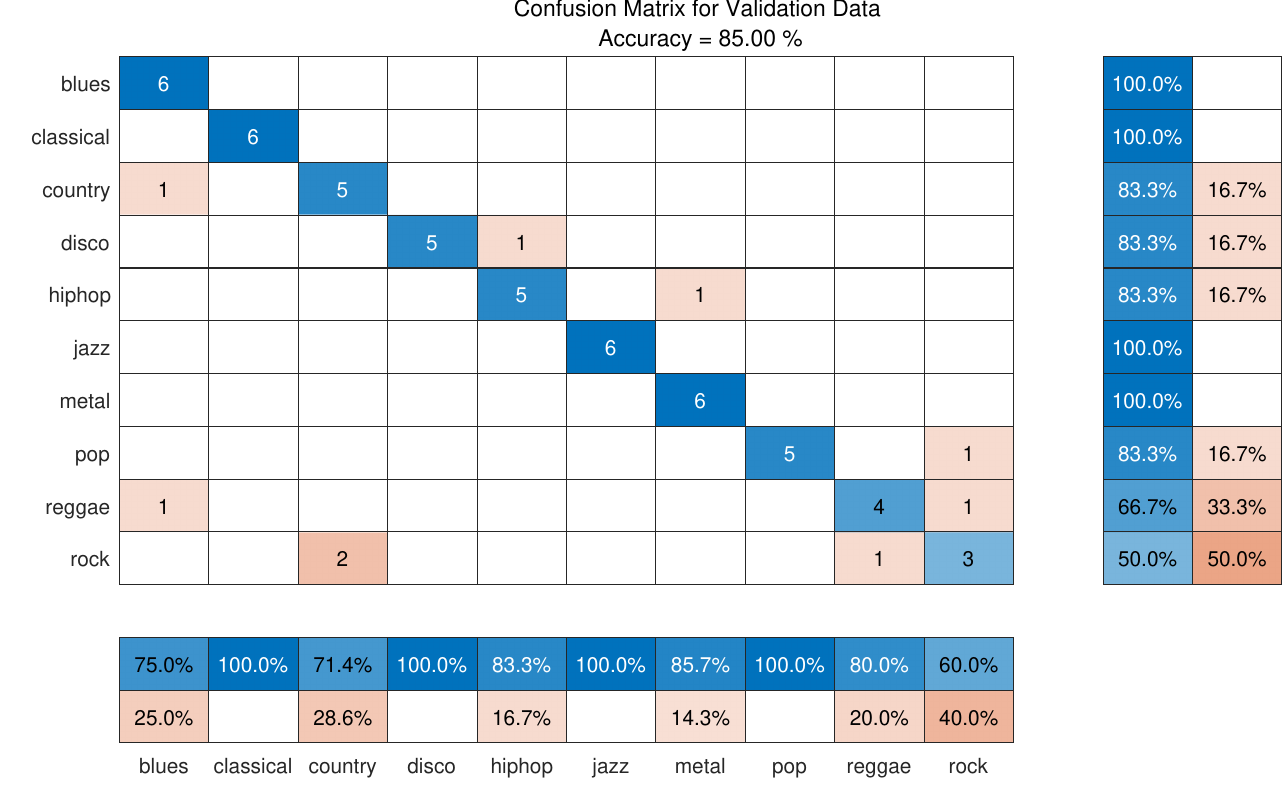}
	\caption{
		\textbf{
			Classification precision of the proposed BAN on Music Genres dataset. 
		}
		For various audio clips, the inferred classification outputs are determined with BAN.
	}
	\label{fig:confusion_matrix}
\end{figure}

To verify that the brain activity of participants and BAN captured enough information to differentiate between music label in the experimental launch, we executed music recognition using a decoding model based on brain and BAN activity. 
As shown in Fig.~\ref{fig:confusion_matrix}, we evaluated the confusion matrix and recognition accuracy (the diagonal elements of the confusion matrix) by analyzing brain response within genre representation region of interest. 
The recognition results varied across music label, with classical music consistently classified accurately (average classification accuracy of 100\%), while rock music showed poor classification performance across participants (40\%). 
Additionally, participants often misclassified reggae as rock (confusion rate of 33.3\%) and rock as country (confusion rate of 28.6\%). 
The confusion matrices derived from activity were highly consistent across participants (Spearman’s correlation coefficient, $ \rho = 0.562 \pm 0.087$; $ p < 0.001 $ for all participant combinations).

\begin{figure}[t]
	\centering
	\includegraphics[width=\linewidth]{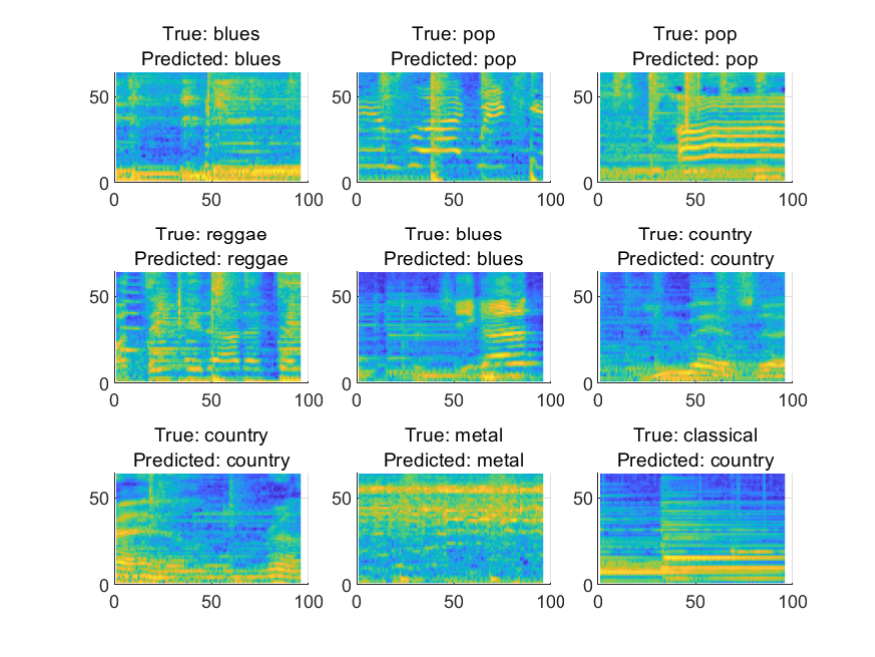}
	\caption{
		\textbf{
			Visualize predictions. 
		}
		View a sample of the input data along with the true and predicted class labels. 
		The $ x $-axis represents time, the $y$-axis represents frequency, and the colormap indicates decibel levels. 
		For several classes, distinct features are clearly visible. 
		For instance, the spectrogram for the country music class displays simple melodies and steady rhythms over time, characteristic of country music. 
		Additionally, it highlights the low-frequency sounds produced by musical instruments typical of this genre.
	}
	\label{fig:predicted_results}
\end{figure}

\begin{figure}[t]
	\centering
	\includegraphics[width=\linewidth]{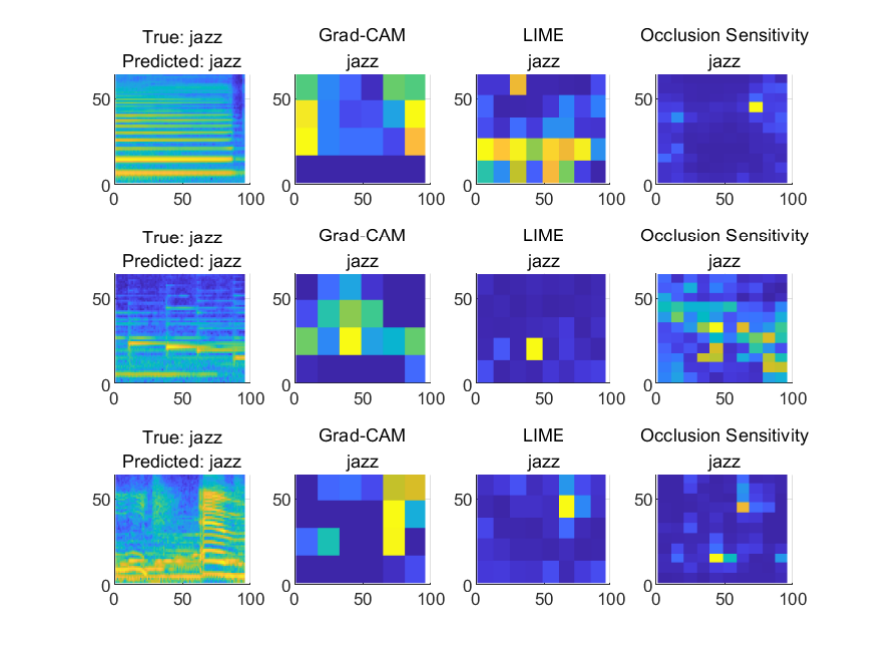}
	\caption{
		\textbf{
			Investigate predictions. 
		}
		Analyze the predictions of the validation Mel spectrograms. For each input, generate Grad-CAM, LIME, and occlusion sensitivity maps for the predicted classes. 
		These techniques take an input image and a class label, producing a map that highlights the regions of the image most important to the score for the specified class. 
		Each visualization method employs a distinct approach that influences the output it generates.
	}
	\label{fig:specific_class}
\end{figure}

As shown in Fig.~\ref{fig:specific_class}, Grad-CAM uses the gradient of the classification score with respect to the BAN features to identify which parts of the input are most critical for classification. 
Regions with high gradients indicate areas where the final score is most influenced by the data. 
LIME approximates BAN's classification behavior using a simpler, more interpretable model, such as a linear model or a regression tree, to assess the importance of input features as a proxy for their relevance to BAN. 
Occlusion sensitivity works by perturbing small parts of the input with an occluding mask, and as the mask moves across the input, the method measures changes in the probability score for a given class.

\begin{figure}[t]
	\centering
	\includegraphics[width=\linewidth]{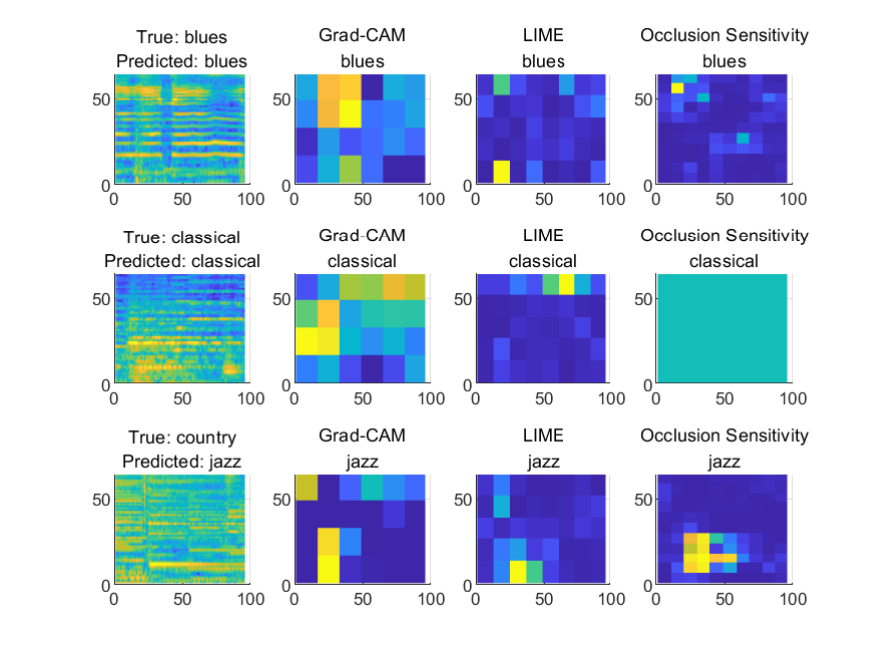}
	\caption{
		\textbf{
			Investigate predictions for specific class. 
		}
		Explore the interpretability maps for spectrograms associated with a specific class.
	}
	\label{fig:investigate_predictions}
\end{figure}

\begin{figure}[t]
	\centering
	\includegraphics[width=\linewidth]{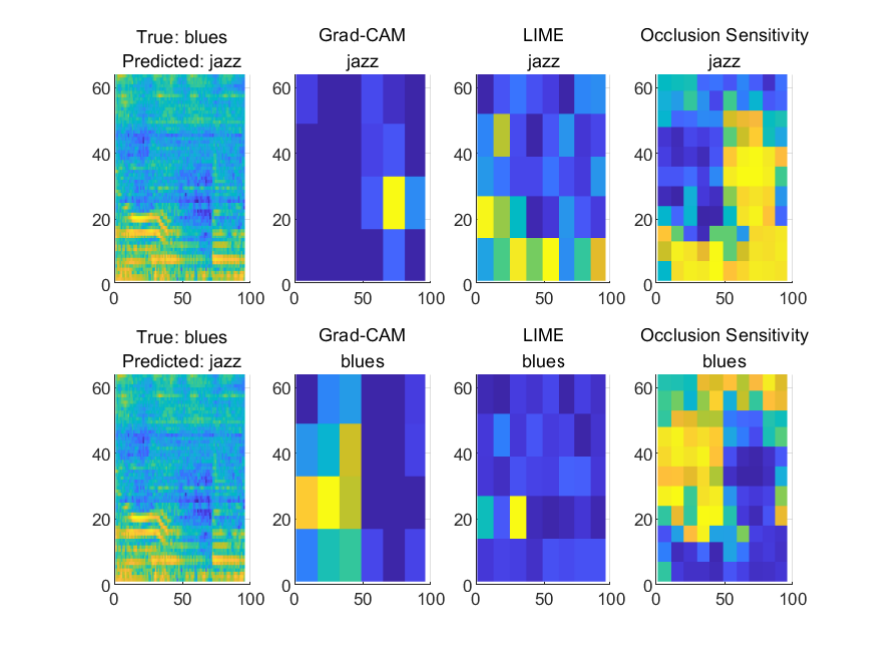}
	\caption{
		\textbf{
			Use the interpretability maps to examine misclassifications. 
		}
		Analyze a spectrogram where the true class is blues, but the predicted class is jazz.
	}
	\label{fig:investigate_misclassifications}
\end{figure}

As shown in Fig.~\ref{fig:investigate_predictions}, spectrograms corresponding to the blues class were identified. 
Interpretability maps were generated and plotted using the input spectrograms and predicted class labels. 
The maps reveal that the network focuses on areas of high intensity and low frequency. 
This result is surprising, as one might expect the network to also consider the high-frequency noise that repeats over time. 
Identifying such patterns is crucial for understanding the features the network relies on to make predictions. 
Similarly, as depicted in Fig.~\ref{fig:investigate_misclassifications}, maps were generated and plotted for both the true class (blues) and the predicted class (jazz).

\subsection{Genre Representation in BAN and Cortex} \label{sec:cortical_organization}

The cortical areas involved in genre representation were evaluated using BAN. 
Significant prediction accuracy was observed in the bilateral STG across all participants ($ p < 0.05 $, FDR corrected). 
To identify brain region that consistently shown music labels, regardless of sample selection, we determined the genre-representing functional region of interest for each participant utilizing a resampling procedure. 
This analysis confirmed significant prediction accuracy in the bilateral STG, and the functional region of interest was utilized as an mask in subsequent analyses.

To evaluate the contribution of each brain voxel to 10 music label, we mapped genre representations on the cortical surface using principal component analysis with BAN weights. 
For each voxel within the region of interest, we extracted the learned BAN weights and applied principal component analysis for dimensionality reduction on the aggregated BAN features. 
The score indicated how the 10 principal components were shown in each brain voxel, while the loading matrix showed the contribution of each principal component to the representation of 10 music genres. 
To visualize the cortical organization of music genres for each participant, we extracted and normalized the principal component analysis scores from their respective voxels. 
This analysis revealed various genre-specific representations within the bilateral STG. 
As shown in TABLE~\ref{tab:prediction_accuracies}, music genres were more distinctly represented in Heschl's sulcus and the lateral STG compared to Heschl's gyrus, the planum temporale, or the lateral sulcus (with the exception of sub-03, who exhibited genre-specific activations in large brain regions, including planum temporale). 
Despite individual variability, a consistent pattern emerged: pop, disco, country, and hip-hop had stronger representations in the LSTG, while blues music showed larger contributions around the HS.


\begin{table}
	
	\centering  
	\caption{The comparison between classical model (Music Class, MFCC) and BAN prediction performance in each anatomical area. 
		LSTG, lateral superior temporal gyrus; HG, Heschl's gyrus; HS, Heschl's sulcus; PT, planum temporale; MFCC, mel-frequency cepstral coefficient}
	\label{tab:prediction_accuracies} 
	
	\resizebox{\linewidth}{!}{ 
		\begin{tabular}{llll} 
			\toprule[1.0pt]
			&    	    Music Class &		  MFCC &    				BAN 			    \\
			\midrule
			L.HG &  	0.148 $\pm$ 0.023 &    0.091 $ \pm $ 0.031 &	 0.102 $ \pm $ 0.029 \\
			L.HS &  	0.219 $\pm$ 0.079 &    0.102 $ \pm $ 0.039 &	 0.171 $ \pm $ 0.063 \\
			L.PT &  	0.127 $\pm$ 0.029 &    0.057 $ \pm $ 0.018 &	 0.117 $ \pm $ 0.039 \\
			L.LSTG &  	0.081 $\pm$ 0.011 &   0.032 $ \pm $ 0.008 &	 0.082 $ \pm $ 0.013 \\
			R.HG &  	0.154 $\pm$ 0.021 &    0.103 $ \pm $ 0.018 &	 0.124 $ \pm $ 0.032 \\
			R.HS &  	0.199 $\pm$ 0.036 &    0.108 $ \pm $ 0.047 &	 0.146 $ \pm $ 0.032 \\
			R.PT &  	0.117 $\pm$ 0.041 &   0.061 $ \pm $ 0.022 &	 0.091 $ \pm $ 0.045 \\
			R.LSTG &  	0.103 $\pm$ 0.018 &    0.048 $ \pm $ 0.011 &	 0.102 $ \pm $ 0.021 \\
			\bottomrule[1.0pt]
		\end{tabular}
	}
	
\end{table}

\section{Conclusion}

Inspired by the auditory processing mechanisms of the human brain, our study introduces a brain-like model tailored for auditory recognition tasks. 
By incorporating neuroanatomical constraints, the model achieves strong recognition performance with enhanced interpretability. 
Additionally, we demonstrate how the cortical model aligns with human auditory recognition of real music and propose a new method for calculating the similarity between model activations and brain responses. 
In addition, the neuroanatomically aligned model offers improved predictions of cortical responses. 
We believe that the proposed BAN will inspire new approaches in ANN interpretability and potentially drive advancements in brain-computer interfaces.



%


\bibliographystyle{IEEEtran}
\bibliography{reference}





\ifCLASSOPTIONcaptionsoff
  \newpage
\fi

\end{document}